# Properties of trapped electromagnetic modes in coupled waveguides

G. Annino[1, @], H. Yashiro[2], M. Cassettari[1], and M. Martinelli[1]

[1] *Istituto per i Processi Chimico-Fisici, CNR, via G. Moruzzi 1, 56124 Pisa, Italy*
[2] *KYOKUGEN Institute, Osaka University, Toyonaka, Osaka, 560-8531, Japan*

(Dated: December 16, 2005)

**Abstract**

The existence of trapped modes in coupled electromagnetic waveguides is experimentally investigated for configurations with different degrees of symmetry supporting hybrid modes. The occurrence of confined solutions in such open geometries is proven to be much more general than demonstrated so far, as predicted by Goldstone and Jaffe [J. Goldstone and R. L. Jaffe, Phys. Rev. B **45**, 14100 (1992)]. The identification of the observed modes is based on the numerical modeling of the relative vector fields. The experimental results evidence as an increasing aperture of the configuration can improve the confinement of the mode instead of generating additional leakage channels. In particular, long-lived trapped modes can be easily obtained in geometries with a relevant degree of aperture. Their resonance parameters can compete with those of standard close cavities working at millimeter wavelengths. The role of the symmetry on the properties of these trapped modes is discussed in detail.

_@ E-mail address: geannino@ipcf.cnr.it_

The analysis of nonpropagating solutions confined in curved or coupled guides was mainly stimulated by the accessibility of the mesoscopic regime in semiconductor devices. In particular, it followed the first experimental works on the so-called quantum wires, where the motion of the electrons becomes ballistic and then dictated by the geometrical constraints (see, for instance, Refs. 1-3, and references therein). On the wave of these measurements, the possibility to trap electrons in open domains that cannot bind classical particles was established by several authors [3-7]. The nature of these bound states appeared from the beginning more general than the quantum mechanic one which led to their discovery, since the time-independent Schrödinger equation for the free electron inside an infinite quantum well coincides with the Helmholtz equation subject to Dirichlet boundary conditions [8]. Similar confinement properties are therefore expected in analogous systems defined by a wave equation, as in acoustic or electromagnetic problems and, as more recently pointed out, in atom or phonon waveguides [9, 10].

In case of electromagnetic systems, long-lived resonance modes in open configurations were investigated in detail in the context of laser physics (see Ref. 11 and references therein). Nevertheless, the existence of a wave confinement (trapped modes) in simple open geometries as curved or coupled waveguides was demonstrated theoretically [12, 13] and experimentally [14-16] only more recently. The analysis of these confined solutions (also termed bound states in the quantum mechanics literature) was limited to transverse electric (TE) modes of waveguides with rectangular cross section, which can be described with the help of a single component vector potential. The relevance of such TE modes is motivated by the fact that they mimic the wave behavior of a corresponding 2-dimensional quantum waveguide. The interest for electromagnetic modes trapped in coupled waveguides was therefore mostly derived, with few exceptions [12, 17], by that for quantum systems, and not considered of real intrinsic relevance. On the other hand, the working principle underlying the wave confinement in coupled rectangular waveguides, namely the availability at their intersection of a volume that is in some sense larger than that accessible to the radiation in the straight waveguide [7], is expected to lead to trapped modes under much more general conditions, as postulated by Goldstone and Jaffe in their basic paper [13].

The aim of the present paper is twofold. First, the prediction of Goldstone and Jaffe will be experimentally verified, on the basis of the observation of trapped modes in geometrical configurations with different symmetry levels involving transverse as well as hybrid modes. The observed modes will be identified by means of a finite-element numerical modeling. Second, long-lived modes, namely high quality factor resonances, will be demonstrated in largely open devices, a property that can prelude to their practical application as resonant element. The relation between the symmetry of the configuration and the properties of these long-lived modes will be discussed in detail. In the following, the mathematical terminology typically employed in the description of coupled waveguides will be freely borrowed, without a real ambition of mathematical rigor. More formal definitions can be found for instance in Refs. [18, 19].

The essential result obtained for a system formed by two identical rectangular waveguides coupled to each other in L, T, right-angle, X, and parallel configuration is that at least a trapped mode exists below the cutoff frequency of the modes propagating along the straight waveguide [5, 7, 13, 17, 20-22]. In the idealized condition of infinite waveguides and perfect conductors, this mode corresponds to an ideal eigenstate of the Laplacian operator subject to Dirichlet boundary conditions. In any practical case the finite size of the system and the finite conductivity of the employed material transform this mode into a resonance with finite linewidth [12, 13]. Perhaps still more remarkable, other trapped modes can exist above the cutoff of the waveguide. These modes are embedded in the continuous spectrum of the operator and in general do not decay due to symmetry reasons [5, 23]. The existence of trapped modes embedded in the continuum has been also demonstrated in particular asymmetric configurations, showing that their projection on the propagating modes vanishes [22]. The departure from the idealized situation transforms again these perfectly trapped modes in resonances with finite linewidth.

The generalization of the results mentioned above will be investigated on different open configurations realized as coupled waveguides, as shown in Figs. 1a-1c at increasing order of symmetry. Fig. 1a shows in particular the right-angle crossing of two different circular waveguides, realized drilling orthogonal holes in a metallic block. In Fig. 1b, an additional waveguide is drilled perpendicularly to those of Fig. 1a. Fig. 1c finally shows the right-angle coupling of a circular waveguide with a parallel-plate waveguide. The excitation of all these configurations is obtained exploiting the coupling of the possible trapped modes with the continuum of modes propagating in the open space, as discussed in [13]. This principle can be practically implemented as shown in Fig. 1d; the intersection of the waveguides is placed close to the side of the block, in correspondence of the excitation waveguide. The radiation penetrates in the intersection region through the narrow waveguide. As far as the frequency of the radiation is below the cutoff of such waveguide an almost complete reflection occurs, unless a trapped mode causes a resonant absorption of the incoming radiation. The reflected wave can be decoupled by the incoming one by means of a circulator, before reaching the detector. More details about a similar setup can be found in [24].

The configuration of Fig. 1a is a variant of the right-angle crossing of rectangular waveguides studied for instance in Refs. [5, 17]. As in that case, the common region of the two waveguides is expected able to trap electromagnetic modes, following the basic principle underlying the wave confinement in these devices. The main difference with respect to the standard configuration is the variable cross-section shape of the region accessible to the radiation, for which no simple transverse modes are expected. The system can be seen as composed by a central region (housing possible trapped modes) open towards the continuum by four waveguides that can radiate away the energy. As a consequence, the cutoffs of the modes propagating along the waveguides represent important frequency markers. In this configuration trapped modes are expected below the minimum cutoff of the two waveguides. Additional confined solutions could exist above such frequency. However, they are expected at frequencies where there is for both waveguides a limited number of propagating modes. These modes represent indeed possible leakage channels for any solution confined at the intersection of the two waveguides. A trapped mode must necessarily have vanishing coupling with each propagating modes. The number of resulting conditions increases with the number of such modes. Therefore, a consistent set of equations is obtained in general only for frequencies allowing a limited number of propagating modes, as shown for instance by Linton and Ratcliffe for two parallel waveguides coupled laterally [22].

The symmetry of the configuration can introduce simple selection rules for the coupling between trapped modes and propagating modes. In the present case, the basic symmetry is that about the axis of each circular waveguide. The role of this symmetry can be discussed introducing the modal indices that label the propagating modes. The first of these indices, commonly indicated as $n \in N_0$, determines the behavior of the mode along the azimuthal angle $\varphi$ about the axis of the waveguide, given by $\cos(n\varphi)$ or $\sin(n\varphi)$. It defines in particular the symmetry of the mode with respect to this axis and is therefore the only relevant for the present analysis.

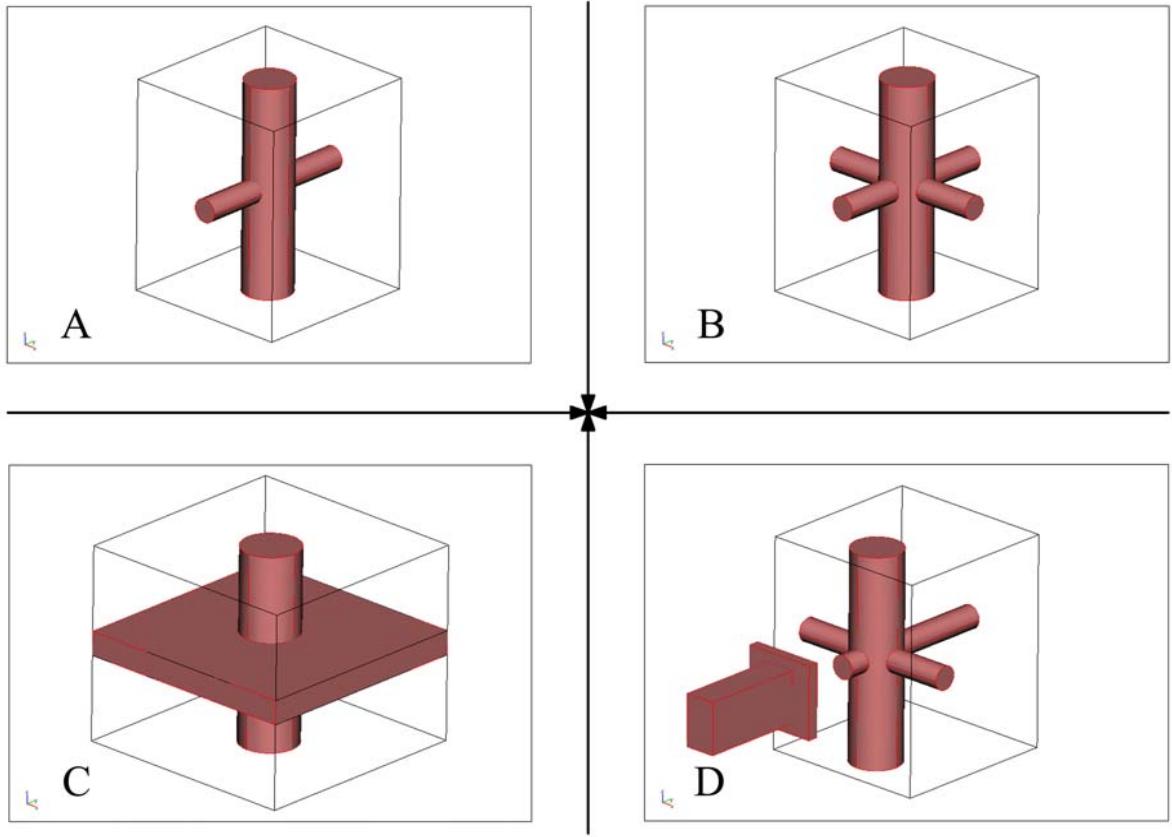

Fig. 1. Coupled-waveguide configurations investigated in the paper. The shaded regions represents the waveguides, the parallelepiped the metallic block in which the waveguides are drilled. The drawings are in scale. (A) Right-angle crossing of two waveguides. (B) Right-angle crossing of three waveguides. The narrow ones have the same diameter. (C) Right-angle crossing of a circular waveguide with a parallel-plate waveguide. (D) Actual working configuration. The intersection of the waveguides is positioned close to a side of the metallic block. The excitation is ensured through a rectangular waveguide in contact with such side.

In case of Fig. 1a, there is a twofold rotational symmetry with respect to the axis of each waveguide, neglecting the presence of the excitation setup. In particular, the configuration has a mirror symmetry plane including the axes of both waveguides. A second mirror symmetry plane is orthogonal to the first one and includes the axis of the large waveguide. The third symmetry plane is orthogonal to the first two and includes the axis of the narrow waveguide. As a consequence of this symmetry, the whole field distribution can be obtained by that in the quarter of the structure comprised between the first two symmetry planes. These planes can be replaced by perfect electric conductor (PEC) or perfect magnetic conductor (PMC), depending on the parity of the different field components with respect to them, according for instance to the analysis in Ref. [25] and references therein. The normal modes with even azimuthal index $n = 2k$, $k \in N_0$, are compatible with homologous boundary conditions on the first two symmetry planes, while the modes with odd azimuthal index $n = 2k+1$ are compatible with alternate boundary conditions on such planes. In the particular case $n = 0$, the $TE_{0m}$ modes are compatible with PEC planes while the $TM_{0m}$ modes are compatible with PMC planes, for any value of the radial modal index $m$. The possible trapped modes of the configuration of Fig. 1a can thus be developed either in terms of odd or even modes of both waveguides.

The practical realization of the configuration of Fig. 1a employed an anticorodal alloy block in which two orthogonal waveguides with $4.65 \pm 0.1$ mm and $2.25 \pm 0.1$ mm diameter were drilled. The block had a square cross section with side of 16 mm, and a height of 20 mm. The distance between the rim of the large waveguide and the closest boundary of the metallic block was $1.85 \pm 0.1$ mm.

The choice of waveguides with sensibly different diameters shows practical and conceptual benefits. First, the cutoffs of the narrow waveguide are at sensibly higher frequencies than those of the large waveguide. As a consequence, the resonance frequencies of the hypothetical trapped modes are expected lower than the cutoffs of the narrow waveguide, as necessary for a proper use of the proposed excitation scheme. In such case, only the propagating modes of the large waveguide are in principle involved in the leakage of the trapped modes. The narrow waveguide can be seen as the perturbation that traps the modes at the intersection of the two waveguides. On the other hand, the essential properties of the trapped modes are expected to be independent on the relative size of the coupled waveguides.

The experimental investigation of the above configuration was performed in the frequency interval ranging from 42 to 92.5 GHz, covered by means of a vector network analyzer as described in Ref. [24]. In such interval the power delivered by the source shows large variations. A resonant absorption in the reflected wave was clearly visible adjusting in a proper way the relative position between the excitation waveguide and the metallic block. The maximum absorption was obtained for incoming radiation with linear polarization orthogonal to the axis of the large waveguide. In order to extract the signal absorbed by the structure from the power fluctuations of the source, a normalization procedure was employed. In particular, the resonant absorption can be eliminated inserting in the large waveguide a quartz tube filled with a high-loss liquid as water. In this manner, the modes trapped at the intersection of the waveguides are suppressed and the reflected signal can be employed to normalize that obtained in the unperturbed system. The typical spectrum resulting from this procedure is shown in Fig. 2. The main resonant absorption at 59.27 GHz is characterized by an unloaded quality factor $Q_0$=1800, determined from the measured one as discussed in Ref. [24] and references therein. Two other peaks are visible around 80 GHz. Their shape is however largely affected by the position of the excitation waveguide. The spikes at frequencies higher than 85 GHz are noise generated by an imperfect normalization procedure, as results comparing the original curves.

The field distributions allowed by the geometries under analysis were modeled by means of a finite-element software (FEMLAB 3.1, Comsol, Se), which solves the Helmholtz equations $\left(\nabla^2 + k^2\right)\begin{Bmatrix}\vec{E}\\\vec{H}\end{Bmatrix} = 0$ for the electric and magnetic fields. In case of perfect conductors, such fields satisfy the conditions $\hat{n} \times \vec{E}\big|_{bound} = 0$ and $\left(\hat{n} \times \left(\nabla \times \vec{H}\right)\right)\big|_{bound} = 0$ on the conducting surfaces, where $\hat{n}$ is the normal to the surface. The symmetry of the configuration can be employed to reduce the volume of analysis and then the numerical effort. In case of Fig. 1a, the twofold symmetry along the axis of the large waveguides and the third symmetry plane discussed earlier allow a reduction of such volume to an octant of the real volume, as shown in the inset of Fig. 2. The effect of the excitation setup on the spectrum of the trapped mode is assumed negligible. The boundary conditions on the surfaces of the reduced volume can be given by PEC or PMC conditions. The obtained resonance modes can be labeled accordingly. In particular, the nomenclature adopted here is based on the parity of the magnetic field along the axis of the large waveguide with respect to the symmetry planes [25]. The modes will be labeled as ABC$_i$, with A replaced by S or N for modes in which the axial magnetic field is even or odd with respect to the median symmetry plane, respectively. In the same manner, B will indicate the parity with respect to the symmetry plane including the axes of the two waveguides, and C the parity with respect to the remaining symmetry plane. Finally, the index *i* is a progressive number that orders for ascending frequencies the resonance modes which satisfy the same boundary conditions.

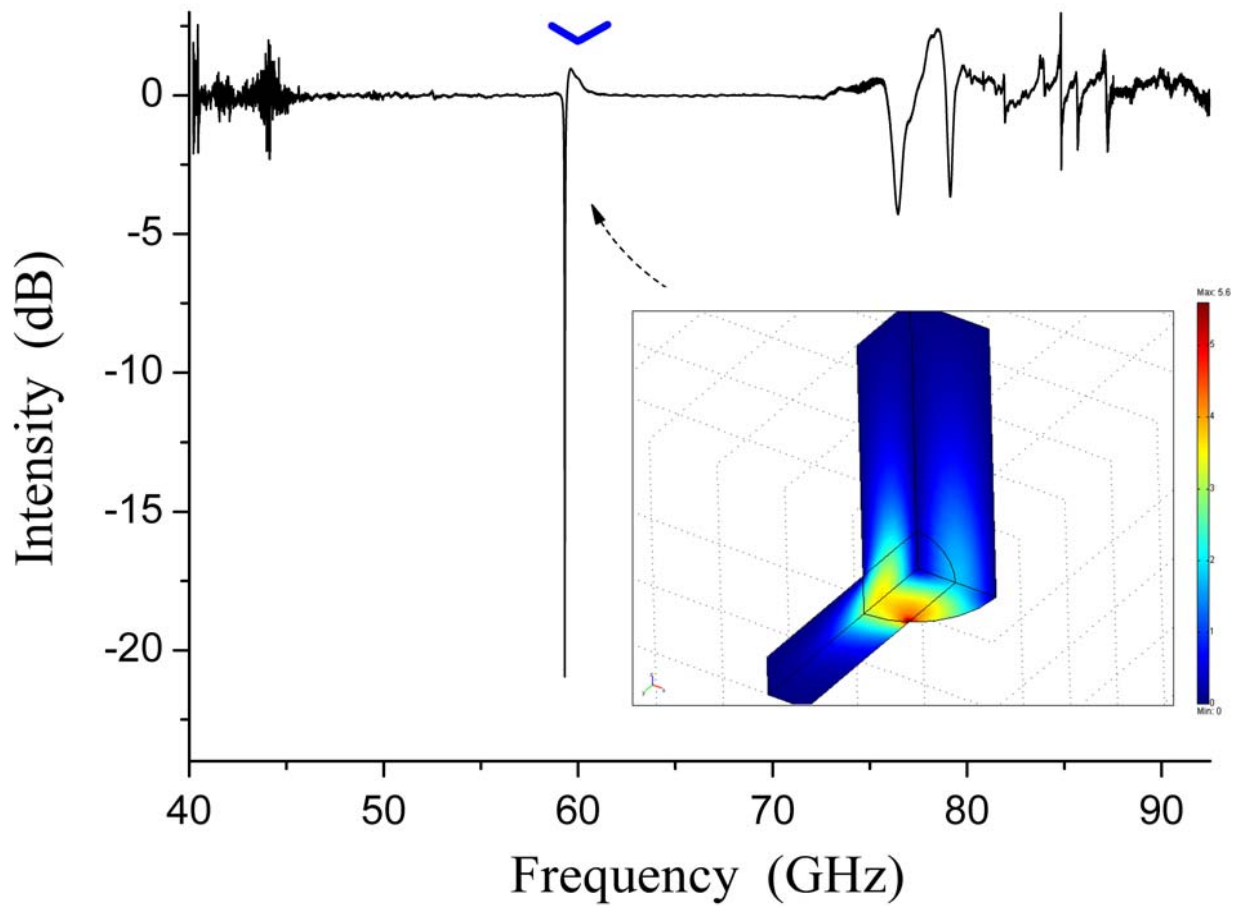

Fig. 2. Normalized absorption spectrum of the configuration shown in Fig. 1a. The V-shaped label indicates the calculated resonance frequency and the relative uncertainty due to the mechanical tolerances. The inset shows the norm of the electric field calculated for the indicated resonance mode, plotted on the surfaces of symmetry. On its right the colors scale, expressed in arbitrary units.

The effect of the open boundaries can be taken into account replacing them with the so-called perfectly matched layer (PML) absorbing boundary conditions. The benefit of this numerical trick is that an open domain can be accurately modeled by means of a finite volume [19]. The modeling returns in this case complex resonance frequencies in which the imaginary component quantifies the loss of energy due to the leakage of radiation.

In case of Fig. 1a the modeling gives one clearly confined mode at 59.96 GHz, which well corresponds to the observed resonance taking into account the uncertainty due to the mechanical tolerances, given by ± 1.6 GHz. Such solution, obtained assuming a PMC median plane and PEC axial symmetry planes, represents a hybrid mode with polarization prevalently orthogonal to the axis of the main waveguide. According to the convention proposed earlier, this mode can be labeled as SSS$_1$. The norm of its electric field on all the symmetry planes is shown in the inset of Fig. 2. Here, the vertex of the V-shaped label indicates the resonance frequency predicted by the modeling, the width of the label the relative uncertainty. The effect of the PMLs on this mode vanishes asymptotically when the layers are positioned at increasing distances from the intersection of the two waveguides. As a consequence, in the limit of infinite waveguides this resonance corresponds to a perfectly trapped mode. The comparison of its frequency with the first cutoffs of the large waveguides, given by 37.78 GHz (TE$_{11}$ mode), 49.35 GHz (TM$_{01}$ mode), and 62.68 GHz (TE$_{21}$ mode) [26], shows that the observed mode is embedded in the continuous spectrum of such waveguide. The lowest cutoff of the narrow waveguide is on the other hand placed at 78.09 GHz. The decay of this trapped mode is prevented by the symmetry reasons discussed earlier, which forbid the coupling with the propagating modes with even modal index. The coupling with the TM$_{01}$ mode vanishes as well, since the boundary conditions required by such particular mode are not compatible with those of the observed SSS$_1$ mode. The frequency interval accessible to the SSS$_1$ mode extends therefore at least to 62.68 GHz, namely to the cutoff of the next even index mode.

The modeling predicts two other hybrid modes with a certain degree of localization, resonating at 37.6 GHz (SSN$_1$) and 49.1 GHz (NNN$_1$). The quality factor of these solutions is of the order of few hundreds. Such figure is limited by the irradiation losses along the axis of the large waveguide, since both solutions are largely distributed along this direction. Although not observable in the employed experimental configuration, these modes present some remarkable features. In particular, the NNN$_1$ mode is well polarized along the large waveguide. On the contrary, the polarization of the SSN$_1$ mode is prevalently orthogonal to this direction. The nature of these solutions can be clarified increasing the length of the coupled waveguides. In such transformation both solutions remain confined around the intersection region; accordingly, their Q factor shows a rapid increase with the size of the configuration. They represent therefore precursors of real trapped modes. The SSN$_1$ solution corresponds in particular to an isolated trapped mode, being below the lowest cutoff of the employed waveguides. The large spreading of these two modes is explained by the closeness of their frequency to the cutoff of the propagating modes compatible with their field distribution. An enlargement of the diameter of the narrow waveguide reduces the resonance frequency of these modes and then improves their confinement. No other confined solutions exist in the investigated frequency interval, according to the numerical modeling. The peaks observed around 80 GHz are most probably induced by the excitation setup. All the predicted solutions are characterized by a field distribution resembling a TE or a TM mode. This feature is reminiscent of the TE or TM nature of the normal modes of the large waveguide, since as discussed the narrow waveguide can be considered as a perturbation.

The configuration of Fig. 1b was investigated following a similar approach. In this case, two nominally identical narrow waveguides were drilled orthogonal to each other and to the large one, both in the same plane. The resulting structure has a fourfold rotational symmetry with respect to the axis of the large waveguide and a twofold symmetry with respect to the narrow waveguides, neglecting the presence of the excitation setup. The higher symmetry allows a further reduction of the volume of analysis, which can be limited to a $\frac{\pi}{4}$ sector about the axis of the large waveguide due to the additional mirror symmetry planes bisecting the two secondary waveguides. Analogously, the higher symmetry introduces stronger selection rules for the coupling with the modes of the main waveguide. Homologous boundary conditions admit now modes with indices $n = 2 \cdot 2k$, while alternate boundary conditions admit modes with $n = 2 \cdot (2k+1)$. The $n = 0$ case can be treated as above. Half of the possible modes of the main waveguide are formally not compatible with the symmetry of the present structure [27]. On the other hand, the onset of the leakage regime through the narrow waveguides remains the same, since it depends only on their diameter. The presence of the further waveguide reduces therefore the number of modes by which the energy confined at the center of the system can radiate away, extending the frequency interval accessible to the trapped modes. The above considerations lead to the conclusion that the additional aperture (and then the new potential leakage channel) reduces the irradiation losses, instead to increase them as it could be supposed at first sight. This rather counterintuitive behavior is peculiar of the wave nature of the investigated phenomena, and explains in particular the existence of non-classical bound states for ballistic electrons in crossed quantum wires.

The configuration of Fig. 1b was realized by means of an anticorodal alloy block in which a waveguide with $4.85 \pm 0.1$ mm diameter is crossed by two nominally identical waveguides with $2.25 \pm 0.1$ mm diameter. The size of the block is the same of the previous case. The distance between the rim of the large waveguide and the closest boundary of the metallic block was $1.85 \pm 0.1$ mm. Following the same procedure employed above, the spectrum of Fig. 3 was obtained. Two resonant absorptions are now clearly visible. The first resonance, centered at 55.55 GHz, is characterized by $Q_0$=1400. The second resonance is centered at 71.14 GHz and characterized by $Q_0$=1100. The excitation of both modes is maximal for radiation with polarization orthogonal to the axis of the main waveguide. The spikes at frequencies above 80 GHz are due to the imperfect normalization procedure. The two broad signals around 75 GHz are strongly dependent on the position of the excitation waveguide. The modeling of this configuration gives two well confined hybrid modes. The first one is a $SSN_1$ mode resonating at $55.4 \pm 1.6$ GHz, the second one a $SSS_1$ mode resonating at $71 \pm 2$ GHz. In both cases the electromagnetic field has polarization prevalently orthogonal to the axis of the main waveguide, in agreement with the experimental observations. The irradiation losses of these modes tend to disappear displacing the PMLs far from the intersection of the three waveguides. The $SSN_1$ mode corresponds to the $SSS_1$ solution of the previous configuration, as verified numerically for a gradual transformation of the geometry of Fig. 1a to that of Fig. 1b.

The norm of the electric field on the symmetry planes is reported in the insets of Fig. 3 for both the observed modes. The V-shaped labels indicate as usual the calculated resonance frequencies and the relative uncertainties. The first relevant cutoffs of the large waveguides are given by 47.32 GHz ($TM_{01}$ mode), 60.09 GHz ($TE_{21}$ mode), 75.39 GHz ($TE_{01}$ mode), whereas the first cutoff of the narrow waveguides is above 100 GHz. The two trapped modes are therefore embedded in the continuum. Their decay through the propagating modes is prevented by the selection rules imposed by the symmetry.

No additional resonances emerge from the modeling, except for a broad $NNN_1$ mode centered around 46.80 GHz. This solution corresponds to the $NNN_1$ mode of the previous case. As expected, it shows characteristics similar to those of its analogous. Finally, the two broad signals observed around 75 GHz are most probably due to the excitation setup.

The last investigated configuration, shown in Fig. 1c, can be considered to some extent a limit case of those of Fig. 1a and Fig. 1b. Here an infinite number of waveguides, all lying on the same plane, are inserted along the all possible directions orthogonal to the main waveguide. The resulting structure possesses a complete rotational invariance about the axis of the circular waveguide (assuming plates with circular cross section), in addition to the mirror symmetry with respect to the median plane. As a consequence of the rotational symmetry, all solutions of the electromagnetic equations can be associated to a modal index $n$. The selection rules require now the conservation of this index. Therefore, the number of modes of the circular waveguide by which the possible trapped modes can radiate away is further reduced.

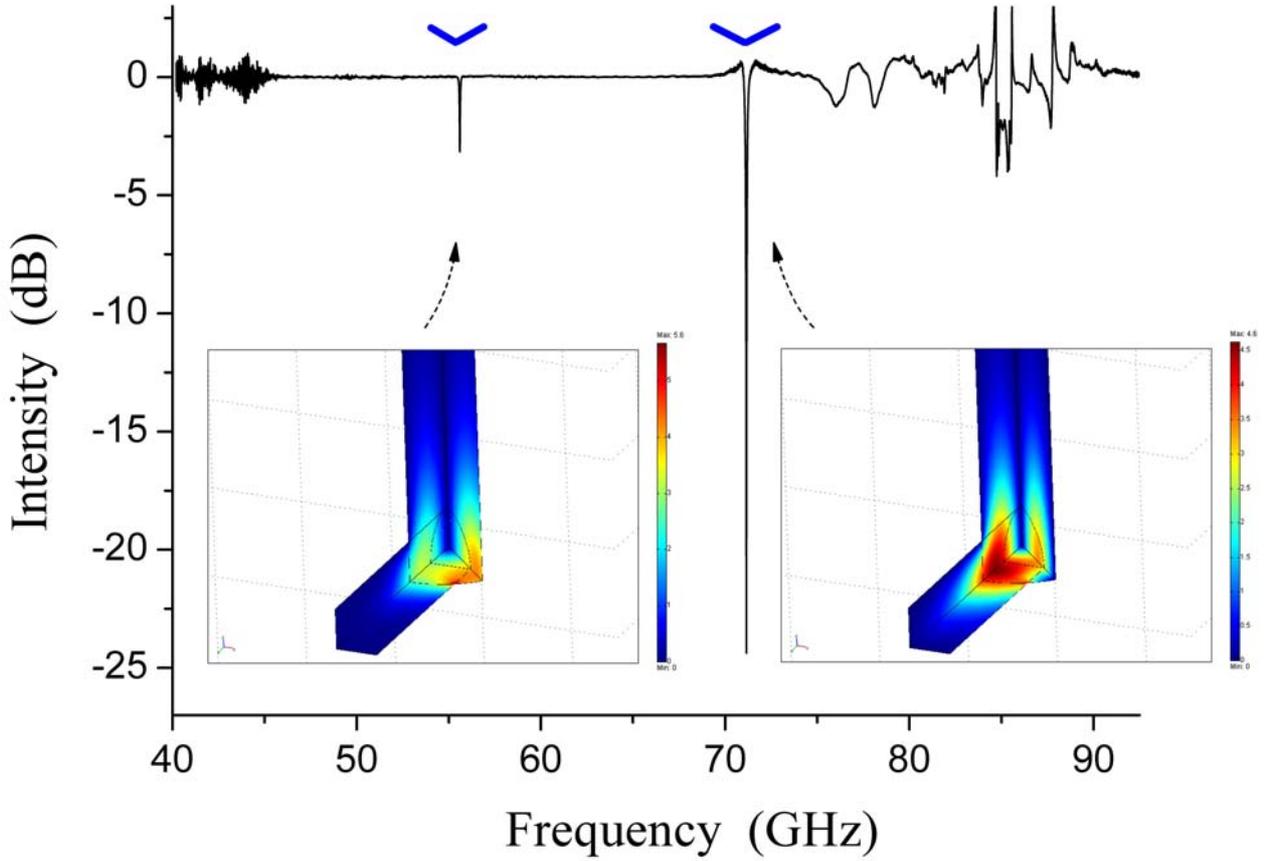

Fig. 3. Normalized absorption spectrum of the configuration shown in Fig. 1b. The V-shaped labels indicate the calculated resonance frequencies and the corresponding uncertainties due to the mechanical tolerances. The insets show the norm of the calculated electric field of the two resonance modes, plotted on the surfaces of symmetry. On the right of each inset the colors scale, expressed in arbitrary units.

The propagation of radiation along the parallel-plate waveguide requires a different analysis, since the modes of this structure cannot be obtained from those of a circular waveguide. In particular, a parallel-plate waveguide supports transverse magnetic (TM) modes without cutoff, as discussed for instance in Refs. [25, 28]. These modes are characterized by axial invariance. The remaining modes have a cutoff that satisfies the condition $v_{cutoff} \geq \frac{c}{2l}$, where $l$ is the distance between the plates and $c$ the speed of light in vacuum. The discussion about the influence of these modes on the spectrum of the resonator is postponed after the experimental and the numerical results.

The practical realization of the geometry of Fig. 1c employed two square anticorodal alloy plates with 6 mm height and 16 mm side, in which a $4.1 \pm 0.05$ mm hole was drilled at $1.4 \pm 0.05$ mm from one of the lateral edges The distance between the plates was fixed to $1.7 \pm 0.05$ mm. The excitation of the resulting configuration was obtained in the usual way, placing the external waveguide on the side of the plate closest to the intersection region. The polarization of the incident radiation was orthogonal to the axis of the circular waveguide, in order to ensure a complete reflection in absence of trapped modes. The absorption spectrum was obtained normalizing the signal acquired with the excitation waveguide in correspondence of the intersection region to that acquired far from it, since in the latter case the radiation is decoupled from the possible trapped modes [24]. The typical result of this procedure is shown in Fig. 4. A first resonance is observed at 67.76 GHz, with a quality factor $Q_0=500$. A second resonance is centered at 81.44 GHz, with $Q_0=2400$. The third resonance is centered at 89.80 GHz, with $Q_0=200$.

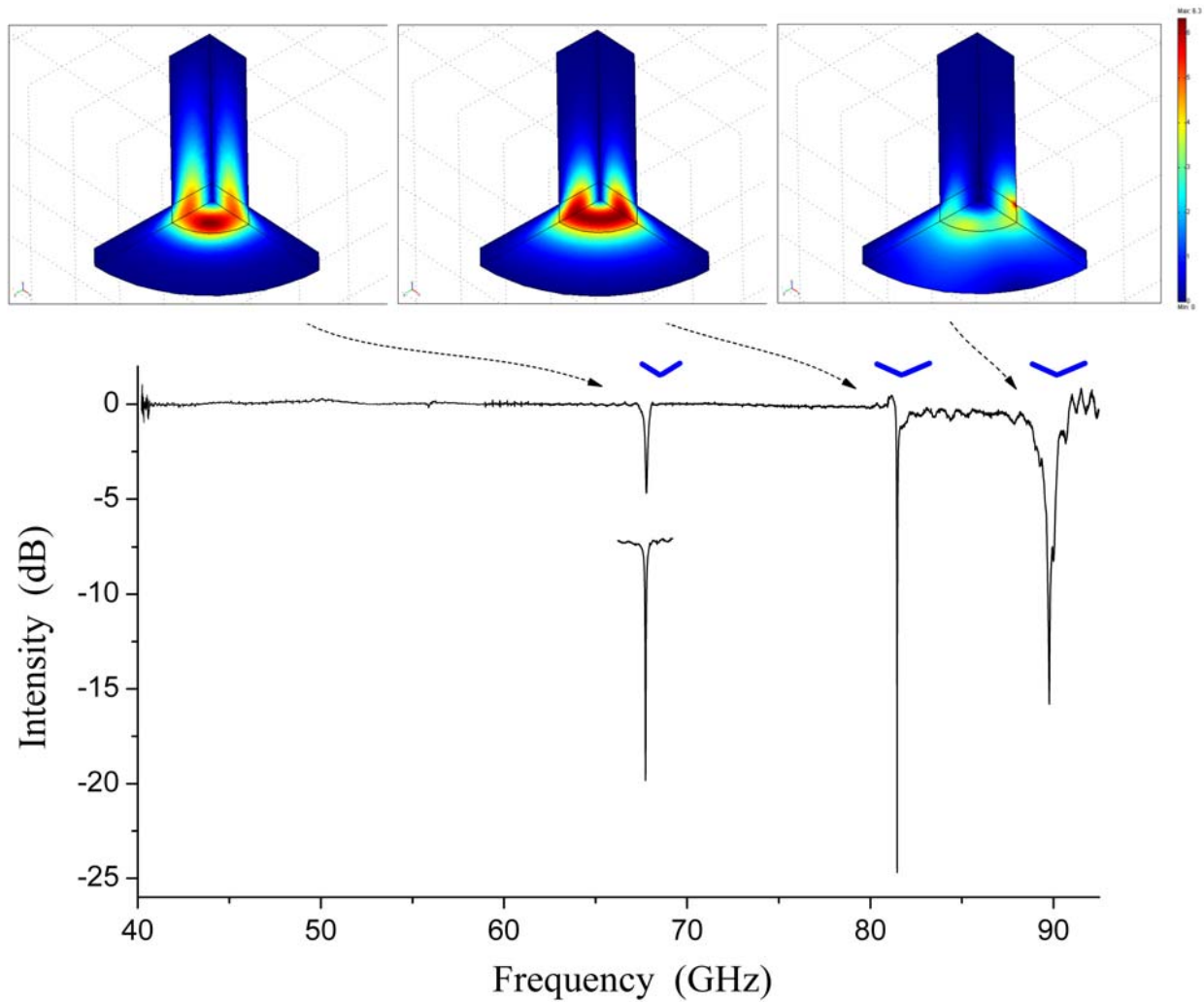

Fig. 4. Normalized absorption spectrum of the configuration shown in Fig. 1c. The lower curve in correspondence of the first resonance represents the same mode, as obtained in a configuration with thicker plates (see text). The V-shaped labels indicate the calculated resonance frequencies and the corresponding uncertainties due to the mechanical tolerances. The insets show the norm of the calculated electric field of the three resonance modes, plotted on the surfaces of symmetry. On the right of the insets the colors scale, expressed in arbitrary units.

In the modeling of the geometry of Fig. 1c, the volume of analysis can be reduced according to the index $n$ of the modal family under investigation. However, an octant of the real structure is compatible with all modal indices, provided that the possible combinations of boundary conditions are considered. According to the numerical modeling, the first resonance of Fig. 4 corresponds to the hybrid $SSS_1$ mode, which is indeed expected in the interval $68.5 \pm 1$ GHz. The norm of the relative electric field on the symmetry planes is reported in the first inset of Fig. 4. The modal index inferred by the field behavior is $n = 2$. The second resonance corresponds to the $SSS_2$ mode, which is expected at $81.8 \pm 1.5$ GHz. The norm of the relative electric field is shown in the second inset of Fig. 4. This mode is characterized by azimuthal invariance ($n = 0$) and by transverse electric fields. Therefore, it corresponds to the well-known $TE_{011}$ mode of the cylindrical resonators [29]. Such mode is indeed always compatible with the geometries having rotational invariance [30]. The last resonance corresponds to the hybrid $SSN_2$ mode, expected at $90.3 \pm 1.5$ GHz. The norm of the electric field on the symmetry planes, shown in the third inset of Fig. 4, evidences a modal index $n = 3$. A gradual transformation of the geometry of Fig. 1b to that of Fig. 1c shows that the $SSS_1$ and $TE_{011}$ modes arise from the $SSN_1$ and $SSS_1$ modes of the previous configuration, respectively.

The resonance frequency of the $SSS_1$, $TE_{011}$, and $SSN_2$ modes can be compared with the lowest cutoff of the modes with identical azimuthal index propagating along the circular waveguide, given by 71.1 GHz ($TE_{21}$ mode), 89.2 GHz ($TE_{01}$ mode), and 97.8 GHz ($TE_{31}$ mode), respectively. The closeness of the $SSS_1$ mode to the cutoff of the related $TE_{21}$ mode explains its wide distribution along the circular waveguide and, in turns, its relatively low quality factor. If the thickness of both plates is increased from 6 mm to 12 mm, the quality factor of this resonance raises indeed appreciably (from 500 to 1200), as evidenced by the lower curve of Fig. 4. On the other hand, the lowest cutoff among all the possible modes propagating along the circular waveguide is well below the frequencies of the observed resonances. Their decay through such axial modes is prevented by the selection rules. An analogous situation characterizes the parallel-plate waveguide, which supports a family of TM modes without cutoff. The decay of the resonances through these modes is forbidden again by symmetry reasons, since their (S) parity with respect to the median plane of the configuration is opposite to the (N) parity of the cutoff-less propagating modes [25, 31]. The observed resonances are therefore embedded in a continuous spectrum of modes propagating along the circular as well as the parallel-plate waveguide. Their irradiation losses vanish asymptotically with the extension of the employed configuration, according to the numerical modeling. The observed resonances represent therefore real trapped modes. No other trapped modes are expected in the frequency interval of interest, excluding a largely distributed $SSN_1$ mode placed around 42.3 GHz. This mode, characterized by azimuthal index $n = 1$, manifests an increasing degree of confinement for configurations with higher axial extension. It can be thus classified as precursor of a trapped mode.

Following the trend evidenced by the previous cases, the enlargement of the aperture in the circular waveguide that transforms the configuration of Fig. 1b to that of Fig. 1c improves the capability of the structure to support trapped modes. The generality of this conclusion extends much beyond the cases discussed here, as confirmed by the numerical modeling in a wide range of conditions.

A circular waveguide coupled to a parallel-plate waveguide resembles a standard cylindrical cavity in which the body of the device is cut in two parts and the plungers are removed. The $TE_{011}$ modes of these two configurations show strong similarities; in a sense, they are the same mode. They allow therefore a direct comparison between the performances of a standard close cavity and those of the open configuration of Fig. 1c. In the latter case, the ohmic losses introduced by the finite conductivity of the bulk anticorodal limit the quality factor of the TE$_{011}$ mode of the investigated configuration to $Q_{0,\Omega}$=5900, according to a perturbative approach in which this quantity follows from the fields obtained assuming perfect conductors. In such approach, the quality factor is given by $Q_{0,\Omega} = \frac{2}{\delta} \frac{\int_{vol} H^2 dV}{\int_{sur} H_t^2 dS}$, where $\delta$ is the skin depth in the conductor and $H_t$ the component of the magnetic field tangential to the conducting surface. A similar calculation applied to an anticorodal standard cavity resonating at the same frequency leads to an optimal merit factor of $Q_{0,\Omega}$=7450. Analogous proportions follow from the calculated density of electromagnetic energy, since the active volumes of the two resonators are comparable. As a consequence, the ideal performances of the open cavity realized coupling a cylindrical waveguide with a parallel-plate waveguide can compete with those of a standard cavity. Analogous results have been obtained in preliminary measurements comparing actual configurations based on the same material and the same level of finishing. High quality factors are also expected for the trapped modes of more general configurations.

The configurations with rotational invariance show the further remarkable feature to be always compatible with transverse modes without azimuthal dependence (where the reference direction is given by the axis of symmetry) [25]. Such modes will be hereafter indicated as TE$_0$ and TM$_0$. The transverse modes are in general describable in terms of a single field component. In case of solutions with modal index $n = 0$, the independent field component can be the azimuthal electric field $E_\varphi$ for the TE$_0$ modes and the azimuthal magnetic field $H_\varphi$ for the TM$_0$ modes. The electric field $E_\varphi$ and the covariant component of the magnetic field $\rho H_\varphi$ satisfy, in the limit of ideal conductors, modified wave equations with Dirichlet ($E_\varphi|_{bound} = 0$) and Neumann ($\hat{n} \cdot \nabla (\rho H_\varphi)|_{bound} \equiv \frac{\partial}{\partial n}(\rho H_\varphi)|_{bound} = 0$) boundary conditions, respectively, where $\rho$ is the radial coordinate [32].

The configurations discussed above were characterized by a relatively high degree of symmetry. Other geometries realized by crossing waveguides with different relative position and shape were investigated numerically, following a reverse path in which the symmetry of the configuration is reduced until its complete removal. The nonsymmetrical configurations obtained from a gradual deformation of symmetric configurations were first considered, monitoring the irradiation losses of the trapped modes. A gradual deformation of the configuration leads in general to a gradual perturbation of the confined solutions. An instructive case is obtained displacing the axis of the narrow waveguide with respect to that of the large waveguide in the configuration of Fig. 1a. During the displacement the resonance frequency of the trapped mode under analysis was kept constant by varying appropriately the diameter of the narrow waveguide. By this way the number of propagating modes by which the trapped mode can decay was kept constant as well. The dissipation in the metallic walls was disregarded assuming a perfect conductor. In such conditions a 0.1 mm displacement, corresponding to the typical geometrical uncertainty of the investigated devices, did not introduce significant irradiation losses on the trapped modes. At 0.4 mm, the asymmetry lowered the merit factor of the SSS$_1$ mode to about 1000 and that of the NNN$_1$ mode to about $2 \cdot 10^4$, while the losses of the SSN$_1$ mode remained unaffected. At 0.8 mm the merit factor of the SSS$_1$ mode was depressed to about 150 and that of the NNN$_1$ mode to about 7000; the losses of the SSN$_1$ mode remained still negligible. This is the typical behavior observed in the other generic configurations investigated here. Under the effect of the deformation the trapped modes are perturbed according to their frequency and to the residual symmetry of the configuration. In particular, the decay of a trapped mode is in general controlled by the symmetry of the configuration, which influences the coupling with the propagating modes, and by its resonance frequency, which defines the number of such modes. The confinement of the isolated trapped modes is insensitive to the symmetry, since there are no decay channels at their frequency. This peculiarity suggests that confined solutions can exist in coupled waveguides under very general conditions.

In conclusion, the existence of trapped modes in coupled electromagnetic waveguides appears as a much more general phenomenon than demonstrated so far, in agreement with the prediction of Goldstone and Jaffe [13]. In particular, hybrid trapped modes exist in a variety of geometrical configurations, as demonstrated by experimental and modeling results. The confinement of such modes can be improved by a larger aperture of the configuration. The improvement is more pronounced when a larger aperture corresponds to a higher degree of symmetry. The symmetry of the configuration plays a fundamental role in the determination of the spectrum of the trapped mode since it controls the coupling with the propagating modes. For generic configurations the perturbation of the symmetry reduces the frequency interval allowed to the trapped modes, generating irradiation losses which increase with the number of modes propagating along the waveguides. The isolated trapped modes are therefore not affected by the level of symmetry; they can thus exist under very general conditions.

The merit factor of the observed modes is remarkably high and can compete with that of close metallic cavities working at millimeter wavelengths. It can be moreover weakly dependent on the perturbation of the structure and then on the mechanical imperfections. A system of coupled waveguides can therefore represent a robust and effective resonant device with a surprising degree of aperture, as in case of the configuration with rotational invariance investigated here or in that recently discussed by Linton and Ratcliffe in the framework of scalar fields [34]. Such simple resonators can find useful applications in any measurement technique requiring a large access to the sample.

In case of geometries having rotational invariance and of modes without azimuthal dependence, the vector electromagnetic problem can be reduced to a scalar problem with either Dirichlet or Neumann boundary conditions. In such case, configurations of coupled electromagnetic waveguides can be employed in principle as model system for the investigation of the wave phenomena in different fields